\documentclass[twocolumn,showpacs,showkeys,aps,superscriptaddress]{revtex4}
\usepackage{graphicx}

\usepackage{color}

\usepackage{bm}
\usepackage{amsmath}
\usepackage{amsfonts}
\usepackage{amssymb}
\usepackage[latin1]{inputenc}
\usepackage{amsthm}

\newcommand{\ket}[1]{|#1\rangle}
\newcommand{\bra}[1]{\langle#1|}

\def\Tr{{\rm Tr}}

\begin{document}

\title{On the concurrence of superpositions of many states}
\author{Seyed Javad Akhtarshenas}
\email{akhtarshenas@phys.ui.ac.ir} \affiliation{Department of
Physics, University of Isfahan, Isfahan, Iran } \affiliation{Quantum
Optics Group, University of Isfahan, Isfahan, Iran}

\date{\today}
\begin{abstract}
In this paper we use the \textit{concurrence vector}, as a measure
of entanglement, and investigate lower and upper  bounds on the
concurrence of a superposition of bipartite states as a function of
the concurrence of the superposed states. We show that the amount of
entanglement quantified by the concurrence vector is exactly the
same as that quantified by \textit{I-concurrence}, so that our
results can be compared to those given in [Phys. Rev. A {\bf 76},
042328 (2007)]. We obtain a tighter lower bound in the case that two
superposed states are orthogonal. We also show  that when the two
superposed states are not necessarily orthogonal, both lower and
bounds are, in general, tighter than the bounds given in terms of
the I-concurrence.  An extension of the results to the case with
more than two states in the superpositions is also given.
\end{abstract}



\maketitle

\section{Introduction}

\label{sec:intro} Quantum entanglement is one of the most
challenging feature of quantum mechanics which has recently
attracted much attention in view of its connection with the theory
of quantum information and computation. It turns out that quantum
entanglement provides a fundamental potential resource for
communication and information processing, therefore, both
characterization and quantification of the entanglement are
important tasks in the theory of quantum information.

Although entanglement is a global property of a state of bipartite
system, originating from the superposition of different states, but
the entanglement of a superposition of pure states cannot be simply
expressed as a function of the entanglement of the individual states
in the superposition. Recently, Linden, Popescu and Smolin
\cite{LindenPRL2006} have raised the following problem: Given a
bipartite quantum state $\ket{\Gamma}$ and a certain decomposition
of it as a superposition of two others
\begin{equation}\label{Gamma}
\ket{\Gamma}=\alpha\ket{\psi}+\beta\ket{\phi},
\end{equation}
what is the relation between the entanglement of $\ket{\Gamma}$ and
those of the two states in the superposition?  They found an upper
bound on the entanglement of $\ket{\Gamma}$ in terms of the
entanglement of $\ket{\psi}$ and $\ket{\phi}$, using the von Neumann
entropy of the reduced state of either of the parties as a measure
of entanglement \cite{BennettPRA1996}.

In order to have motivation on this problem, it is worth considering
two simple examples of a two-qubit system \cite{LindenPRL2006}. Let
us first consider a state of two-qubit system as
$\ket{\gamma}=\frac{1}{\sqrt{2}}\ket{0}\ket{0}+\frac{1}{\sqrt{2}}\ket{1}\ket{1}$.
As we know, each term in the superposition is unentangled, yet the
superposition  is a maximally entangled state of the two-qubit
system. On the other hand, consider
$\ket{\gamma^\prime}=\frac{1}{\sqrt{2}}\ket{\Phi^+}+\frac{1}{\sqrt{2}}\ket{\Phi^-}$,
where
$\ket{\Phi^{\pm}}=\frac{1}{\sqrt{2}}\ket{00}+\frac{1}{\sqrt{2}}\ket{11}$.
In this case, it is evident that each of the terms in the
superposition is maximally entangled, yet the superposition itself
is unentangled. It follows therefore that the entanglement of a
superposition may be very different from the entanglement of the
superposed states.

Several authors have generalized the results of \cite{LindenPRL2006}
to include different measures of entanglement
\cite{YuPRA2007,OuPRA2007,NisetPRA2007,GourPRA2007}, superpositions
of more than two states \cite{GourPRA2008}, and superpositions of
multipartite states \cite{CavalcantiPRA2007,SongPRA2007}.

The aim of this paper is to use the notion of \textit{concurrence
vector} \cite{AkhtarshenasJPA2005},  as a measure of entanglement,
and provide tight lower and upper bounds on the entanglement of
superpositions of more than two pure states of a bipartite system.
Investigation of the entanglement of the superposition of two
states, by using generalized concurrence as a measure of
entanglement, are given recently in \cite{YuPRA2007,NisetPRA2007}.
We shall see that our definition of the generalized concurrence,
based on the concurrence vector,  is completely equivalent to the
generalized concurrence introduced by Rungta \textit{et al.}
\cite{RungtaPRA2001}, known as \textit{I-concurrence}, so that we
can directly compare our results with those of Ref.
\cite{NisetPRA2007}. We show that when the two superposed states are
orthogonal, our lower bound is tighter than that given in
\cite{NisetPRA2007}, and also in the case that the two superposed
states are not necessarily orthogonal, both lower and upper bounds
are, in general, tighter than those given \cite{NisetPRA2007}. An
extension of the problem to include superpositions of more than two
states is also presented. In the case that the states being
superposed are biorthogonal, we obtain an exact solution for the
concurrence of superpositions. In other situations, where the states
in the superposition are one-sided orthogonal, orthogonal, and
arbitrary,  we present tight lower and upper bounds on the
concurrence of superposition.

The paper is organized as follows. First, in section II, we give a
brief review on the concurrence vector, and show that this
definition is equivalent to the so-called I-concurrence. In this
section we also provide some definitions and useful lemmas, which
will be used repeatedly in the subsequent sections. Section III is
devoted to provide bounds on the concurrence of superpositions of
two terms. In section IV, we extend our results and investigate the
concurrence of superpositions of more than two states. The paper is
concluded in section V, with a brief conclusion.

\section{Concurrence vector}
We begin by introducing the generalized concurrence, based on the
definition of the concurrence vector.  The generalized concurrence
for a bipartite pure state $\left|\psi\right>\in{\mathbb
C}^{N_1}\otimes {\mathbb C}^{N_2}$ is given by
\cite{AkhtarshenasJPA2005}
\begin{equation}\label{CPsi}
C(\psi)=\sqrt{\sum_{\alpha} \sum_{\beta}|C_{\alpha\beta}(\psi)|^2},
\end{equation}
where $C_{\alpha\beta}(\psi)$, with
${\alpha=1,\cdots,N_1(N_1-1)/2}$, ${\beta=1,\cdots,N_2(N_2-1)/2}$
are the components of the concurrence vector ${\bf C}(\psi)$,
defined by
\begin{equation}\label{CAlphaBetaPsi}
C_{\alpha\beta}(\psi)=\bra{\psi}S_{\alpha\beta}\ket{\psi^\ast}.
\end{equation}
Here $S_{\alpha\beta}=L_{\alpha}\otimes L_{\beta}$,  where
$L_{\alpha}$ and $L_{\beta}$ are generators of $SO(N_1)$ and
$SO(N_2)$, respectively.

For further use, we extend the above definition of the concurrence
of a pure state  to the concurrence of two pure states. Let
$\ket{\psi}$ and $\ket{\phi}$ be two pure states of ${\mathbb
C}^{N_1}\otimes {\mathbb C}^{N_2}$. We define $C(\psi,\phi)$ as the
concurrence of $\ket{\psi}$ and $\ket{\phi}$ by
\begin{equation}\label{CPsiPhi}
C(\psi,\phi)=\sqrt{\sum_{\alpha}
\sum_{\beta}|C_{\alpha\beta}(\psi,\phi)|^2},
\end{equation}
where $C_{\alpha\beta}(\psi,\phi)$ are the components of the
concurrence vector ${\bf C}(\psi,\phi)$, defined by
\begin{eqnarray}\label{CAlphaBetaPsiPhi}\nonumber
C_{\alpha\beta}(\psi,\phi)&=&\bra{\psi}S_{\alpha\beta}\ket{\phi^\ast}
\\ & = &
\bra{\phi}S_{\alpha\beta}\ket{\psi^\ast}=C_{\alpha\beta}(\phi,\psi).
\end{eqnarray}
The second line indicates that this definition is symmetric in its
arguments, i.e. ${\bf C}(\psi,\phi)={\bf C}(\phi,\psi)$. Notice that
according to the above definition, we have ${\bf C}(\psi,\psi)={\bf
C}(\psi)$.

Before starting our study, it is worth making observation that the
definition (\ref{CPsi}) for the concurrence is equivalent to the
definition of  the I-concurrence given by Rungta \textit{et al.}
\cite{RungtaPRA2001}. To do so, note that in Eq. (\ref{CPsi}) we
need to calculate the expectation value of the term $\sum_{\alpha}
\sum_{\beta}S_{\alpha\beta}(\ket{\psi^\ast}\bra{\psi^\ast})S_{\alpha\beta}$.
Motivated by this, we define the superoperator $\mathcal{S}$ such
that its action on an arbitrary operator $\sigma$ is given by
\begin{equation} \label{superoperator0}
\mathcal{S}(\sigma)=\sum_{\alpha}\sum_{\beta}S_{\alpha\beta}(\sigma)^{T}S_{\alpha\beta}.
\end{equation}
Now, invoking the definition of $S_{\alpha\beta}$, we find that Eq.
(\ref{superoperator0}) can be written as
\begin{equation} \label{superoperator}
\mathcal{S}(\sigma)=\Tr(\sigma)I_A\otimes I_B-\sigma_A\otimes
I_B-I_A\otimes\sigma_B+\sigma.
\end{equation}
This equation resembles the definition of the universal invertor
superoperator, introduced in Ref. \cite{RungtaPRA2001}. Using this,
we can rewrite the concurrence of Eq. (\ref{CPsi}) as
\begin{eqnarray}
C(\psi)&=&\sqrt{\sum_{\alpha}
\sum_{\beta}\bra{\psi}S_{\alpha\beta}(\ket{\psi^\ast}\bra{\psi^\ast})S_{\alpha\beta}\ket{\psi}}
\\ \nonumber
&=&\sqrt{\bra{\psi}\mathcal{S}(\ket{\psi}\bra{\psi})\ket{\psi}}
=\sqrt{2\left(1-\Tr[\rho_A^2]\right)},
\end{eqnarray}
which is exactly the same as the I-concurrence introduced by Rungta
\textit{et al.} \cite{RungtaPRA2001}.

Now we give below the  definition of biorthogonal and one-sided
orthogonal states.

\textit{Definition 1: Biorthogonal and one-sided orthogonal states.}
Two bipartite pure states $\ket{\psi}$ and $\ket{\phi}$ are called
biorthogonal if satisfy the following conditions
\begin{equation}\label{BiorthogonalStates1}
\Tr_{A}[\rho^A_\psi\rho^A_\phi]=0  \Longleftrightarrow
\bra{e_i^\psi} e_j^\phi\rangle=0,
\end{equation}
and
\begin{equation}\label{BiorthogonalStates2}
\Tr_{B}[\rho^B_\psi\rho^B_\phi]=0  \Longleftrightarrow
\bra{f_i^\psi} f_j^\phi\rangle=0,
\end{equation}
where
$\rho^A_\psi=\Tr_B[\ket{\psi}\bra{\psi}]=\sum_i(\lambda_i^\psi)^2\ket{e_i^\psi}\bra{e_i^\psi}$
and
$\rho^B_\psi=\Tr_A[\ket{\psi}\bra{\psi}]=\sum_i(\lambda_i^\psi)^2\ket{f_i^\psi}\bra{f_i^\psi}$
are the reduced density matrices of $\ket{\psi}$ on the $A$ and $B$
subsystems, respectively,  with $\ket{e_i^\psi}$ and
$\ket{f_i^\psi}$ as the local Schmidt  basis of $\ket{\psi}$, with
the corresponding Schmidt numbers $\lambda_i^\psi$. Similar
definitions are  hold for $\rho^A_\phi$, $\rho^B_\phi$,
$\ket{e_j^\phi}$ and $\ket{f_j^\phi}$.

On the other hand,  we say that  $\ket{\psi}$ and $\ket{\phi}$ are
one-sided orthogonal states  if they satisfy one of the conditions
(\ref{BiorthogonalStates1}) and (\ref{BiorthogonalStates2}). In the
following, without loss of generally, we assume that one-sided
orthogonal states satisfy the first condition
(\ref{BiorthogonalStates1}), but not necessarily the condition
(\ref{BiorthogonalStates2}).

The following lemma hold for the concurrence vectors of a pair of
biorthogonal or one-sided orthogonal states, and will be used in the
following sections.

\textit{Lemma 1.} Let $\ket{\psi}$ and $\ket{\phi}$ be either
biorthogonal or one-sided orthogonal states. Then the inner product
of any pair of the concurrence vectors ${\bf C}(\psi)$, ${\bf
C}(\phi)$ and ${\bf C}(\psi,\phi)$ are zero.

\textit{Proof.}\label{Lemma1Proof1}
 Using Eqs (\ref{CAlphaBetaPsi}), (\ref{CAlphaBetaPsiPhi}) and (\ref{superoperator}) we get
\begin{eqnarray}\nonumber\label{Lemma1Proof1}
{\bf C}(\psi)\cdot {\bf C}(\phi)&=&
2(\bra{\psi}\phi\rangle)^2-\Tr_A[(\Tr_B[\ket{\phi}\bra{\psi}])^2]
\\  &-&\Tr_B[(\Tr_A[\ket{\phi}\bra{\psi}])^2],
\end{eqnarray}
and
\begin{equation}\label{Lemma1Proof2}
{\bf C}(\psi)\cdot {\bf C}(\psi,\phi)=
2\bra{\psi}\phi\rangle-2\Tr_A[(\Tr_B[\ket{\phi}\bra{\psi}])\rho^A_\psi]
\end{equation}
and a similar relation for ${\bf C}(\phi)\cdot {\bf C}(\psi,\phi)$.
Now by writing $\ket{\psi}$ and $\ket{\phi}$ in their Schmidt forms,
it is straightforward to check that every term in the right-hand
side of the Eqs. (\ref{Lemma1Proof1}) and (\ref{Lemma1Proof2}) can
be written as a sum of terms including
$\bra{e_i^\psi}e_j^\phi\rangle\bra{f_k^\psi}f_l^\phi\rangle$. But it
follows from condition (\ref{BiorthogonalStates1}) or
(\ref{BiorthogonalStates2}) that in these cases
$\bra{e_i^\psi}e_j^\phi\rangle=0$ or
$\bra{f_i^\psi}f_j^\phi\rangle=0$, which completes the proof.

The following lemma is a generalization of the above lemma to the
case of more than two biorthogonal or one-sided orthogonal states.

\textit{Lemma 2.} Let $\{\ket{\psi_i}\}$ denotes a set of either
biorthogonal or one-sided orthogonal states. Then the set of
concurrence vectors $\{{\bf C}(\psi_i,\psi_j)\}$  are orthogonal,
i.e. $ {\bf C}(\psi_i,\psi_j)\cdot {\bf
C}(\psi_{i^\prime},\psi_{j^\prime})=0$ unless ${\bf
C}(\psi_i,\psi_j)={\bf C}(\psi_{i^\prime},\psi_{j^\prime})$.

\textit{Proof.} The proof is similar to that given for lemma 1.

\textit{Lemma 3.} Let $\ket{\psi}$ and $\ket{\phi}$ be two pure
states of the bipartite system. Then for the concurrence
$C(\psi,\phi)$ we have the following relations.
\begin{enumerate}
\item If $\ket{\psi}$ and $\ket{\phi}$ are biorthogonal, then
$C(\psi,\phi)=1$.
\item If $\ket{\psi}$ and $\ket{\phi}$ are one-sided orthogonal,  then
$C(\psi,\phi)\le 1$.
\item If $\ket{\psi}$ and $\ket{\phi}$ are orthogonal, then
$C(\psi,\phi)\le 1$.
\item If $\ket{\psi}$ and $\ket{\phi}$ are arbitrary states, then
$C(\psi,\phi)\le \sqrt{1+|\bra{\psi}\phi\rangle|^2}$.
\end{enumerate}

\textit{Proof.} Using Eqs (\ref{CPsiPhi}) and (\ref{superoperator}),
we find
\begin{equation}
C^2(\psi,\phi)=1+|\bra{\psi}\phi\rangle|^2-\Tr_A[\rho^A_\psi
\rho^A_\phi]-\Tr_B[\rho^B_\psi \rho^B_\phi].
\end{equation}
Next, by using the conditions on each class, we obtain the results
of the lemma.

Armed with these lemmas, we are now in a position to study the
bounds on the concurrence of superpositions. We first consider the
case that we have superpositions of two states.

\section{Superpositions of two states}
To begin with, let us consider the more general case where the two
component states in the superposition (\ref{Gamma}) are arbitrary,
not necessarily orthogonal, and therefore the superposition is not
normalized. If we define
$\ket{\Gamma^\prime}=\ket{\Gamma}/\|\Gamma\|$ as the normalized
version of $\ket{\Gamma}$, then we can obtain bounds for the
concurrence of the normalized version of the superposition. To do
so, we plug  Eq. (\ref{Gamma}) into Eq. (\ref{CAlphaBetaPsi}), and
by putting the result into the definition of concurrence, given in
Eq. (\ref{CPsi}), we get the following relation for the concurrence
of  $\ket{\Gamma^\prime}$
\begin{eqnarray}\label{CGammaPrime}\nonumber
\|\Gamma\|^2C(\Gamma^\prime)&=&\{\sum_{\alpha}\sum_{\beta}
\mid(\alpha^\ast)^2C_{\alpha\beta}(\psi)+(\beta^\ast)^2C_{\alpha\beta}(\phi)
\\
&+&2(\alpha^\ast\beta^\ast)C_{\alpha\beta}(\psi,\phi)\mid^2\}^{1/2},
\end{eqnarray}
where $C_{\alpha\beta}(\psi,\phi)$ are defined in Eq.
(\ref{CAlphaBetaPsiPhi}). This relation is general, irrespective of
the orthogonality properties of $\ket{\psi}$ and $\ket{\phi}$. In
the following subsections we will restrict our attention to some
special cases.

\subsection{Biorthogonal states}
First we consider the more restrictive case where the two pure
states $\ket{\psi}$ and $\ket{\phi}$ are biorthogonal states, that
is, they satisfy both conditions given in Eqs.
(\ref{BiorthogonalStates1}) and (\ref{BiorthogonalStates2}). In this
case we can find an exact expression for the concurrence of
superposition.

\textit{Theorem 1: Biorthogonal states.} When $\ket{\psi}$ and
$\ket{\phi}$ are biorthogonal states, then the  concurrence of the
superposition $\ket{\Gamma}=\alpha\ket{\psi}+\beta\ket{\phi}$, with
$|\alpha|^2+|\beta|^2=1$, is given by
\begin{equation} \label{Theorem1Biorthogonal}
C(\Gamma)=\sqrt{
|\alpha|^4C^2(\psi)+|\beta|^4C^2(\phi)+4|\alpha\beta|^2}.
\end{equation}
\textit{Proof.} In this case, according to the lemma 1, the inner
product of the concurrence vectors ${\bf C}(\psi)$, ${\bf C}(\phi)$
and ${\bf C}(\psi,\phi)$ are zero, therefore Eq. (\ref{CGammaPrime})
reduces to
\begin{equation}\label{Theorem1BiorthogonalProof}
C(\Gamma)=\sqrt{
|\alpha|^4C^2(\psi)+|\beta|^4C^2(\phi)+4|\alpha\beta|^2C^2(\psi,\phi)}.
\end{equation}
But recall that $C(\psi,\phi)=1$,  if $\ket{\psi}$ and $\ket{\phi}$
are biorthogonal states (lemma 3). Using this we get Eq.
(\ref{Theorem1Biorthogonal}).

\textit{Remark 1.} If $\ket{\psi}$ and $\ket{\phi}$ are two
biorthogonal states of a two-qubit system, then it requires that
$C(\psi)=C(\phi)=0$, therefore Eq. (\ref{Theorem1Biorthogonal})
reduces to $C(\Gamma)=2|\alpha\beta|$.

\subsection{One-sided orthogonal states}
Now we consider the case of $\ket{\psi}$ and $\ket{\phi}$ being
one-sided orthogonal but not necessarily biorthogonal, that is, they
satisfy condition (\ref{BiorthogonalStates1}).  In this case we can
find the lower and upper bounds for the concurrence of
superposition.

\textit{Theorem 2: One-sided orthogonal states.} When $\ket{\psi}$
and $\ket{\phi}$ are one-sided orthogonal states, then the
concurrence of the superposition
$\ket{\Gamma}=\alpha\ket{\psi}+\beta\ket{\phi}$, with
$|\alpha|^2+|\beta|^2=1$, satisfy
\begin{equation}\label{Theorem2OneSidedOrthogonalUpper}
C(\Gamma)\le\sqrt{|\alpha|^4C^2(\psi)+|\beta|^4C^2(\phi)+4|\alpha\beta|^2},
\end{equation}
and
\begin{equation}\label{Theorem2OneSidedOrthogonalLower}
C(\Gamma)\ge\sqrt{ |\alpha|^4C^2(\psi)+|\beta|^4C^2(\phi)}.
\end{equation}
\textit{Proof.} In this case, again, according to lemma 1  the inner
product of the concurrence vectors ${\bf C}(\psi)$, ${\bf C}(\phi)$
and ${\bf C}(\psi,\phi)$ are zero, therefore Eq.
(\ref{Theorem1BiorthogonalProof}) is satisfied. But from lemma 3 we
have $0\le C^2(\psi,\phi)\le 1$, where completes the proof.

\textit{Remark 2.}  If $\ket{\psi}$ and $\ket{\phi}$ are two
one-sided orthogonal states of a two-qubit system, then they are
necessarily biorthogonal, so again $C(\psi)=C(\phi)=0$, and we find
$C(\Gamma)=2|\alpha\beta|$.

\subsection{Orthogonal states}
Now we assume that the two component states of the superposition
(\ref{Gamma}) are orthogonal, i.e. $\bra{\psi}\phi\rangle=0$, but
not necessarily  one-sided orthogonal.   In this case we find the
following bounds for the concurrence of superposition.

\textit{Theorem 3: Orthogonal states.} Let $\ket{\psi}$ and
$\ket{\phi}$ be two orthogonal states of a general bipartite system.
Then the concurrence of the superposition
$\ket{\Gamma}=\alpha\ket{\psi}+\beta\ket{\phi}$, with
$|\alpha|^2+|\beta|^2=1$, satisfies
\begin{equation}\label{Theorem3OrthogonalUpper}
C(\Gamma)\le |\alpha|^2C(\psi)+|\beta|^2C(\phi)+2|\alpha\beta|,
\end{equation}
and
\begin{equation}\label{Theorem3OrthogonalLower}
C(\Gamma)\ge
\left||\alpha|^2C(\psi)-|\beta|^2C(\phi)\right|-2|\alpha\beta|.
\end{equation}

\textit{proof}: Successive application of the Minkowski inequality
\cite{Minkowski}
\begin{equation}\label{Minkowski}
\left[\sum_{i=1}^{n}|x_i+y_i|^p\right]^{1/p}\le
\left[\sum_{i=1}^{n}|x_i|^p\right]^{1/p}+\left[\sum_{i=1}^{n}|y_i|^p\right]^{1/p},
\end{equation}
with $p>1$, to Eq. (\ref{CGammaPrime}), together with the definition
of the concurrence given in Eq. (\ref{CPsi}), lead to
\begin{equation}\label{Theorem3OrthogonalUpperProof}
C(\Gamma)  \le|\alpha|^2C(\psi)+|\beta|^2C(\phi)
+2|\alpha\beta|C(\psi,\phi).
\end{equation}
Now using the fact that $C(\psi,\phi)\le 1$ (lemma 3),  directly
leads to the upper bound (\ref{Theorem3OrthogonalUpper}).

Next in order to proof the lower bound
(\ref{Theorem3OrthogonalLower}), we make use of the inverse
Minkowski inequality \cite{Minkowski}
\begin{eqnarray}\label{InverseMinkowski}\nonumber
\left[\sum_{i=1}^{n}|x_i+y_i|^p\right]^{1/p}&\ge &
\left|\left[\sum_{i=1}^{n}|x_i|^p\right]^{1/p}-\left[\sum_{i=1}^{n}|y_i|^p\right]^{1/p}\right|
\\
&\ge &
\left[\sum_{i=1}^{n}|x_i|^p\right]^{1/p}-\left[\sum_{i=1}^{n}|y_i|^p\right]^{1/p}
\end{eqnarray}
and apply it to Eq. (\ref{CGammaPrime}), two times. First, by using
the second line of Eq. (\ref{InverseMinkowski}), we separate the
first two terms of Eq. (\ref{CGammaPrime}) from the third one, and
second we use the first line of Eq. (\ref{InverseMinkowski}) and
separate between the first two terms. We get therefore
\begin{equation}\label{Theorem3OrthogonalLowerProof}
C(\Gamma) \ge \left||\alpha|^2C(\psi)-|\beta|^2C(\phi)\right|
-2|\alpha\beta|C(\psi,\phi).
\end{equation}
Again, by using the fact that $C(\psi,\phi)\le 1$, for orthogonal
states, we deduce the advertised inequality
(\ref{Theorem3OrthogonalLower}).

\textit{Remark 3.} If $\ket{\psi}$ and $\ket{\phi}$ are two
orthogonal states of a two-qubit system, then $C(\psi,\phi)\le
\sqrt{1-\delta^2}$, where $\delta=\max\{C(\psi),C(\phi)\}$
\cite{NisetPRA2007}. Applying this to Eqs.
(\ref{Theorem3OrthogonalUpperProof}) and
(\ref{Theorem3OrthogonalLowerProof}), we obtain  tighter  bounds for
the two-qubit case \cite{NisetPRA2007}.

It is worth to mention, however,  that although the upper bound
(\ref{Theorem3OrthogonalUpper}) is exactly the same as the
corresponding upper bound given by Niset \textit{et al.}
\cite{NisetPRA2007}, the lower bound (\ref{Theorem3OrthogonalLower})
is tighter than that they have obtained. The difference is, indeed,
between the third term of Eq. (\ref{Theorem3OrthogonalLower}), in
the sense that the authors of \cite{NisetPRA2007} obtained
$2|\alpha\beta|(1+\delta)$, with
$\delta=\min(|\frac{\alpha}{\beta}|C(\psi_1),|\frac{\beta}{\alpha}|C(\psi_2))$,
instead of $2|\alpha\beta|$. It is interesting to note that they
have suspected that if a more appropriate expression exists, it
would probably have  the correction factor $\delta$ equal to zero.

\subsection{Arbitrary states} We
consider the more general case where the two component states in the
superposition (\ref{Gamma}) are not orthogonal and therefore the
superposition is not normalized. In this case we find the following
bounds for the concurrence of the normalized version
$\ket{\Gamma^\prime}$.

\textit{Theorem 4: Arbitrary states.} Let $\ket{\psi}$ and
$\ket{\phi}$ be two arbitrary states of a general bipartite system.
The concurrence of the normalized state
$\ket{\Gamma^\prime}=\ket{\Gamma}/\parallel\Gamma\parallel$, with
$\ket{\Gamma}=\alpha\ket{\psi}+\beta\ket{\phi}$ and
$|\alpha|^2+|\beta|^2=1$, satisfies
\begin{widetext}
\begin{equation}\label{Theorem4ArbitraryUpper}
\|\Gamma\|^2C(\Gamma^\prime) \le
\min\left\{\begin{array}{l}|\alpha|^2C(\psi)+|\beta|^2C(\phi)+2|\alpha\beta|\sqrt{1+|\bra{\psi}\phi\rangle|^2
},
\\
|\alpha|^2C(\psi)+|\beta^2+2\alpha\beta\bra{\phi}\psi\rangle|C(\phi)+2|\alpha\beta|\sqrt{1-|\bra{\psi}\phi\rangle|^2
}, \\
|\alpha^2+2\alpha\beta\bra{\psi}\phi\rangle|C(\psi)+|\beta|^2C(\phi)
+2|\alpha\beta|\sqrt{1-|\bra{\psi}\phi\rangle|^2
},\end{array}\right.
\end{equation}
and
\begin{equation}\label{Theorem4ArbitraryLower}
\|\Gamma\|^2C(\Gamma^\prime)\ge
\max\left\{\begin{array}{l}\left||\alpha|^2C(\psi)-|\beta|^2C(\phi)\right|-2|\alpha\beta|\sqrt{1+|\bra{\psi}\phi\rangle|^2
},\vspace{1mm}
\\
\left||\alpha|^2C(\psi)-|\beta^2+2\alpha\beta\bra{\phi}\psi\rangle|C(\phi)\right|
-2|\alpha\beta|\sqrt{1-|\bra{\psi}\phi\rangle|^2 },
 \vspace{1mm} \\
\left||\alpha^2+2\alpha\beta\bra{\psi}\phi\rangle|C(\psi)-|\beta|^2C(\phi)\right|-2|\alpha\beta|\sqrt{1-|\bra{\psi}\phi\rangle|^2.
}\end{array}\right.
\end{equation}
\end{widetext}
Note that when the two states $\ket{\psi}$ and $\ket{\phi}$ are
orthogonal, then Eqs. (\ref{Theorem4ArbitraryUpper}) and
(\ref{Theorem4ArbitraryLower}) reduce to Eqs.
(\ref{Theorem3OrthogonalUpper}) and (\ref{Theorem3OrthogonalLower}),
respectively.

\textit{Proof:} For the upper bound (\ref{Theorem4ArbitraryUpper}),
we should  proof that the left-hand side of Eq.
(\ref{Theorem4ArbitraryUpper}) is less that each term in the
right-hand side. For the first line, we proceed exactly the same as
that we done in the proof of the upper bound of theorem 3, but here
we use the fact that $C^2(\psi,\phi)\le
1+|\bra{\psi}\phi\rangle|^2$. Now to reach the second line of Eq.
(\ref{Theorem4ArbitraryUpper}), we write
$\ket{\phi}=\bra{\psi}\phi\rangle\;
\ket{\psi}+\sqrt{1-|\bra{\psi}\phi\rangle|^2}\ket{\psi_{\perp}}$,
where $\ket{\psi_\perp}$ is a vector orthogonal to $\ket{\psi}$.
Putting this into Eq. (\ref{CAlphaBetaPsiPhi}), we find
$C_{\alpha\beta}(\psi,\phi)=\bra{\psi}\phi\rangle^\ast
C_{\alpha\beta}(\psi)+\sqrt{1-|\bra{\psi}\phi\rangle|^2}C_{\alpha\beta}(\psi,\psi_\perp)$,
where can be used in Eq. (\ref{CGammaPrime}), getting
\begin{eqnarray}\nonumber
\|\Gamma\|^2C(\Gamma^\prime) &=&\{\sum_{\alpha}\sum_{\beta}
\mid\left((\alpha^\ast)^2
+2\alpha^\ast\beta^\ast\bra{\psi}\phi\rangle^\ast\right)
C_{\alpha\beta}(\psi) \\
 &+&(\beta^\ast)^2C_{\alpha\beta}(\phi)
\\ \nonumber
&+&2(\alpha^\ast\beta^\ast)\sqrt{1-|\bra{\psi}\phi\rangle|^2}C_{\alpha\beta}(\psi,\psi_\perp)\mid^2\}^{1/2}.
\end{eqnarray}
Next, by successive application of the Minkowski inequality
(\ref{Minkowski}), followed by the fact that $C(\psi,\psi_\perp)\le
1$, we reach to the second line of the inequality
(\ref{Theorem4ArbitraryUpper}).  Alternatively, if we first expand
$\ket{\psi}$ in terms of $\ket{\phi}$ and $\ket{\phi_\perp}$, we get
the third line of the inequality (\ref{Theorem4ArbitraryUpper}).
This completes the proof for the upper bound
(\ref{Theorem4ArbitraryUpper}).

In a similar way, but using the inverse Minkowski inequality instead
of the Minkowski inequality, one can obtain the lower bound
(\ref{Theorem4ArbitraryLower}).

\textit{Remark 4.} If $\ket{\psi}$ and $\ket{\phi}$ are two
arbitrary states of a two-qubit  system, then $C(\psi,\phi)\le
\sqrt{1-\delta^2}$, where
$\delta=\max\{|C(\psi)-|\bra{\psi}\phi\rangle||,|C(\phi)-|\bra{\psi}\phi\rangle||\}$
\cite{NisetPRA2007}. Using this in the process of the above proof,
one can obtain tighter bounds for the two-qubit case
\cite{NisetPRA2007}.

It is worth to mention that  the bounds obtained in Eqs.
(\ref{Theorem4ArbitraryUpper}) and (\ref{Theorem4ArbitraryLower})
are, in general, tighter than the corresponding bounds given by
Niset \textit{et al.} \cite{NisetPRA2007}. In fact, the authors of
\cite{NisetPRA2007} reached to the first line of relation
(\ref{Theorem4ArbitraryUpper}) for the upper bound which is, in
general, weaker than our upper bound. On the other hand, for the
lower bound, they have obtained a term such as the first line of Eq.
(\ref{Theorem4ArbitraryLower}), but with
$2|\alpha\beta|\sqrt{1+|\bra{\psi}\phi\rangle|^2+\delta }$,
$\delta=\min(|\frac{\alpha}{\beta}|C(\psi_1),|\frac{\beta}{\alpha}|C(\psi_2))$,
instead of $2|\alpha\beta|\sqrt{1+|\bra{\psi}\phi\rangle|^2}$.

In order to illustrate the bounds more intuitionally, let us
consider a simple example. Let $\ket{\psi}$ and $\ket{\phi}$ be
states of a ${\mathbb C}^{3}\otimes {\mathbb C}^{3}$ space, defined
by
\begin{eqnarray}
\ket{\psi}&=&\frac{1}{\sqrt{2}}\ket{00}+\frac{1}{2}\ket{11}+\frac{1}{2}\ket{22}, \\
\ket{\phi}&=&\frac{1}{\sqrt{2}}\ket{00}-\frac{1}{2}\ket{11}+\frac{1}{2}\ket{22}.
\end{eqnarray}
Evidently, these states are not orthogonal,
$\bra{\psi}\phi\rangle=1/2$, and that they have equal concurrence,
$C(\psi)=C(\phi)=\sqrt{5}/2$. In Fig. 1, we  illustrate the upper
and lower bounds of the superposition (\ref{Gamma}), with $\alpha=x$
and $\beta=-\sqrt{1-x^2}$. As it is evident from the figure, our
bounds give stronger constraints than those derived in Ref.
\cite{NisetPRA2007}.

\begin{figure}
\centering
\includegraphics[width=8.5 cm]{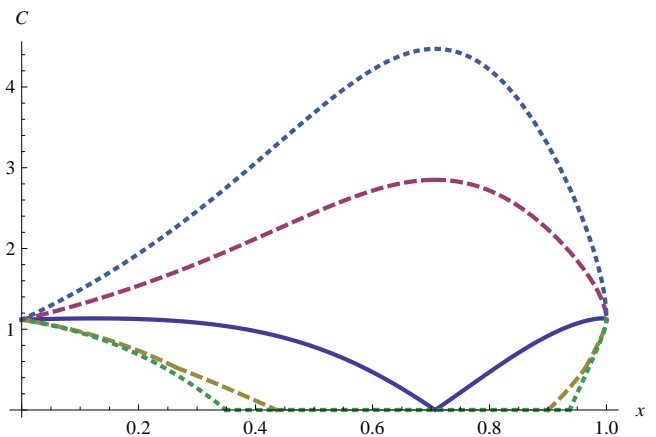}
\caption{(Color online) Concurrence of the normalized version of the
superposition (\ref{Gamma}),
 with
$\ket{\psi}$ and $\ket{\phi}$ as defined in the text. The solid line
is the exact value of $C(\Gamma^\prime)$, while the two dashed lines
correspond to the upper and lower bounds derived from Eqs.
(\ref{Theorem4ArbitraryUpper}) and (\ref{Theorem4ArbitraryLower}),
and the two dotted lines correspond to the upper and lower bounds
given in Ref. \cite{NisetPRA2007}. } \label{Fig1}
\end{figure}

\section{Superpositions of more than two states}
In this section we consider a state of an arbitrary bipartite system
constructed as a superposition of $m$ pure states
\begin{equation}\label{GammaM}
\ket{\Gamma}=\sum_{i=1}^{m}\gamma_i\ket{\psi_i}.
\end{equation}
We  first concern our attention to  the more general case where the
states $\{\ket{\psi_i}\}$ in the superposition (\ref{GammaM}) are
not orthogonal and therefore the superposition is not normalized. We
define  $\ket{\Gamma^\prime}=\ket{\Gamma}/\|\Gamma\|$ as the
normalized version of $\ket{\Gamma}$, then we shall obtain bounds on
the concurrence of the normalized version of the superposition.

To do so, we plug  Eq. (\ref{GammaM}) into Eq.
(\ref{CAlphaBetaPsi}), and by putting the result into the definition
of the concurrence given in Eq. (\ref{CPsi}), we get the following
relation for the concurrence of  $\ket{\Gamma^\prime}$
\begin{eqnarray}\label{CGammaPrimeM}\nonumber
\|\Gamma\|^2C(\Gamma^\prime)&=&\{\sum_{\alpha}\sum_{\beta} \mid
\sum_{i=1}^{m}(\gamma_i^\ast)^2C_{\alpha\beta}(\psi_i)
\\
&+&2\sum_{i<j}^{m}(\gamma_i^\ast\gamma_j^\ast)C_{\alpha\beta}(\psi_i,\psi_j)\mid^2\}^{1/2}.
\end{eqnarray}
In the following subsections we will consider, separately, some
special cases.

\subsection{Biorthogonal states}
First we consider the more restrictive case that the set of pure
states $\{\ket{\psi_i}\}$ are biorthogonal states, that is, they
satisfy conditions (\ref{BiorthogonalStates1}) and
(\ref{BiorthogonalStates2}). In this case we will find an exact
expression for the concurrence of a superposition.

\textit{Theorem 5: Biorthogonal states.} When $\{\ket{\psi_i}\}$ are
biorthogonal states, then the concurrence of the superposition
$\ket{\Gamma}=\sum_{i=1}^{m}\gamma_i\ket{\psi_i}$, with
$\sum_{i=1}^{m}|\gamma_i|^2=1$, is given by
\begin{equation} \label{Theorem1BiorthogonalM}
C(\Gamma)=\sqrt{
\sum_{i=1}^m|\gamma_i|^4C^2(\psi_i)+4\sum_{i<j}^{m}|\gamma_i\gamma_j|^2}.
\end{equation}
\textit{Proof.} In this case, according to the lemma 2, the inner
product of two different concurrence vectors are zero, i.e. ${\bf
C}(\psi_i,\psi_j)\cdot{\bf C}(\psi_{i^\prime},\psi_{j^\prime})=0$
unless  ${\bf C}(\psi_i,\psi_j)={\bf
C}(\psi_{i^\prime},\psi_{j^\prime})$. Therefore we have
\begin{equation} \label{Theorem5BiorthogonalProofM}
C(\Gamma)=\sqrt{
\sum_{i=1}^m|\gamma_i|^4C^2(\psi_i)+4\sum_{i<j}^{m}|\gamma_i\gamma_j|^2C^2(\psi_i,\psi_j)}.
\end{equation}
But recall that $C(\psi_i,\psi_j)=1$ if $\ket{\psi_i}$ and
$\ket{\psi_j}$ are  biorthogonal states. Using this we get Eq.
(\ref{Theorem1BiorthogonalM}).

\subsection{One-sided orthogonal states}
Now we consider the case that the set of pure states
$\{\ket{\psi_i}\}$ are one-sided orthogonal states, that is, they
satisfy the condition (\ref{BiorthogonalStates1}).  In this case we
can find the lower and upper bounds for the concurrence of
superposition.

\textit{Theorem 6: One-sided orthogonal states.} When
$\{\ket{\psi_i}\}$ are one-sided orthogonal states, then the
concurrence of the superposition
$\ket{\Gamma}=\sum_{i=1}^{m}\gamma_i\ket{\psi_i}$, with
$\sum_{i=1}^{m}|\gamma_i|^2=1$, satisfies
\begin{equation}\label{Theorem2OneSidedOrthogonalUpper}
C(\Gamma)\le\sqrt{
\sum_{i=1}^m|\gamma_i|^4C^2(\psi_i)+4\sum_{i<j}^{m}|\gamma_i\gamma_j|^2},
\end{equation}
and
\begin{equation}\label{Theorem2OneSidedOrthogonalLower}
C(\Gamma)\ge\sqrt{ \sum_{i=1}^m|\gamma_i|^4C^2(\psi_i)}.
\end{equation}

\textit{Proof.} In this case again we have that ${\bf
C}(\psi_i,\psi_j)\cdot{\bf C}(\psi_{i^\prime},\psi_{j^\prime})=0$
unless  ${\bf C}(\psi_i,\psi_j)={\bf
C}(\psi_{i^\prime},\psi_{j^\prime})$, therefore Eq.
(\ref{Theorem5BiorthogonalProofM}) is satisfied. But from lemma 3 we
have $0\le C^2(\psi_i,\psi_j)\le 1$, where completes the proof.

\subsection{Orthogonal states}
In this subsection we assume that the set $\{\ket{\psi_i}\}$ are
orthogonal. We find the following bounds for the concurrence of
superposition.

\textit{Theorem 7: Orthogonal states.} Let the set
$\{\ket{\psi_i}\}$ be orthogonal states of a general bipartite
system. Then the concurrence of the superposition
$\ket{\Gamma}=\sum_{i=1}^{m}\gamma_i\ket{\psi_i}$, with
$\sum_{i=1}^{m}|\gamma_i|^2=1$, satisfies

\begin{equation}\label{Theorem7OrthogonalUpperM}
 C(\Gamma) \le
\sum_{i=1}^m|\gamma_i|^2C(\psi_i)+2\sum_{i<j}^{m}|\gamma_i\gamma_j|,
\end{equation}
and
\begin{equation}\label{Theorem7OrthogonalLowerM}
C(\Gamma) \ge  2\Delta -\sum_{i=1}^{m}|\gamma_i|^2C(\psi_i)
-2{\sum_{i<j}^{m}}|\gamma_i\gamma_j|,
\end{equation}
where $\Delta$ is defined by
\begin{equation}\label{Delta}
\Delta=\max\{|\gamma_i|^2C(\psi_i)\}, \qquad i=1,\cdots,m.
\end{equation}

\textit{Proof.} Applying the Minkowski inequality (\ref{Minkowski}),
repeatedly,  to Eq. (\ref{CGammaPrimeM}), followed by the fact that
in this case we have $C(\psi_i,\psi_j)\le 1$, we get the upper bound
(\ref{Theorem7OrthogonalUpperM}).

Next in order to proof the lower bound
(\ref{Theorem7OrthogonalLowerM}), we apply the second line of the
inverse Minkowski inequality (\ref{InverseMinkowski}) to Eq.
(\ref{CGammaPrimeM}) and separate between the first $m$ terms and
the rest of the right-hand side. We get therefore
\begin{eqnarray}\label{Theorem7OrthogonalLowerMProof1}
C(\Gamma)&\ge&\sqrt{\sum_{\alpha}\sum_{\beta} \mid
\sum_{i=1}^{m}(\gamma_i^\ast)^2C_{\alpha\beta}(\psi_i)\mid^2}
\\ \nonumber
&-&2\sqrt{\sum_{\alpha}\sum_{\beta}\mid\sum_{i<j}^{m}(\gamma_i^\ast\gamma_j^\ast)C_{\alpha\beta}(\psi_i,\psi_j)\mid^2}.
\end{eqnarray}
Now by successive application of the inverse Minkowski inequality
and the Minkowski inequality  to the first and the second line of
the above equation, respectively, we get
\begin{equation}\label{Theorem7OrthogonalLowerMProof1}
C(\Gamma)\ge  2\Delta
-\sum_{i=1}^{m}|\gamma_i|^2C(\psi_i)-2{\sum_{i<j}^{m}}|\gamma_i\gamma_j|C(\psi_i,\psi_j).
\end{equation}
Again making use of the fact that $C(\psi_i,\psi_j)\le 1$, for
orthogonal states, we get the lower bound
(\ref{Theorem7OrthogonalLowerM}).

\subsection{Arbitrary states}
In this subsection we generalize the previous theorems to include
the general situation of the superposed states  being arbitrary. We
state the following theorem.

\textit{Theorem 8: Arbitrary states.} Let
$\{\ket{\psi_i}\}_{i=1}^{m}$ be a set of arbitrary normalized, but
not necessarily orthogonal,  pure states of a general bipartite
system. The concurrence of the normalized state
$\ket{\Gamma^\prime}=\ket{\Gamma}/\parallel \Gamma \parallel$, with
$\ket{\Gamma}=\sum_{i=1}^{m}\gamma_i\ket{\psi_i}$,
$\sum_{i=1}^{m}|\gamma_i|^2=1$, satisfies
\begin{widetext}
\begin{equation}\label{Theorem8ArbitraryUpperM}
 \|\Gamma\|^2C(\Gamma^\prime) \le
\min\left\{\begin{array}{l}\sum_{i=1}^m|\gamma_i|^2C(\psi_i)+2\sum_{i<j}^{m}|\gamma_i\gamma_j|\sqrt{1+|\bra{\psi_i}\psi_j\rangle|^2
} \; ,
\\ \\ \min_{\bf P}\{\sum_{k\in\{k<l\}_{-}}|\gamma_k^2
+2\sum_{l\in\{k<l\}_{-}}\gamma_k\gamma_l\bra{\psi_k}\psi_l\rangle|
C(\psi_k) \\
+\sum_{s\in\{r<s\}_{+}}|\gamma_s^2
+2\sum_{r\in\{r<s\}_{+}}\gamma_r\gamma_s\bra{\psi_s}\psi_r\rangle|
C(\psi_s)
\\+\sum_{\overline{i}}^{\prime}|\gamma_{\overline{i}}|^2
C(\psi_{\overline{i}})+2\sum_{i<j}^{m}|\gamma_i\gamma_j|\sqrt{1-|\bra{\psi_i}\psi_j\rangle|^2}\},\end{array}\right.
\end{equation}
and
\begin{equation}\label{Theorem8ArbitraryLowerM}
\|\Gamma\|^2C(\Gamma^\prime) \ge \max\left\{\begin{array}{l}2\Delta
-\sum_{i=1}^{m}|\gamma_i|^2C(\psi_i)
-2{\sum_{i<j}^{m}}|\gamma_i\gamma_j|\sqrt{1+|\bra{\psi_i}\psi_j\rangle|^2
},  \\ \\
\max_{\bf P}\{2\Delta^{\prime} -\sum_{k\in\{k<l\}_{-}}|\gamma_k^2
+2\sum_{l\in\{k<l\}_{-}}\gamma_k\gamma_l\bra{\psi_k}\psi_l\rangle|
C(\psi_k) \\
-\sum_{s\in\{r<s\}_{+}}|\gamma_s^2
+2\sum_{r\in\{r<s\}_{+}}\gamma_r\gamma_s\bra{\psi_s}\psi_r\rangle|
C(\psi_s)
\\-\sum_{\overline{i}}^{\prime}|\gamma_{\overline{i}}|^2
C(\psi_{\overline{i}})-2\sum_{i<j}^{m}|\gamma_i\gamma_j|\sqrt{1-|\bra{\psi_i}\psi_j\rangle|^2}\},\end{array}\right.
\end{equation}
\end{widetext}
where $\min_{{\bf P}}$ ($\max_{{\bf P}}$) denotes minimum (maximum)
over all possible partitions of the set of pairs $\{i<j\}$, with
$1\le i< j \le m$, into distinct subsets $\{k<l\}_{-}$ and
$\{r<s\}_{+}$. The number of such possible partitions is
$2^{m(m-1)/2}$. Also  $\Delta$ is defined in Eq. (\ref{Delta}), and
$\Delta^\prime$ is given by
\begin{equation}
\Delta^{\prime}=\max\left\{\begin{array}{ll} |\gamma_k^2
+2\sum_{l\in\{k<l\}_{-}}\gamma_k\gamma_l\bra{\psi_k}\psi_l\rangle|
C(\psi_k),  \\ |\gamma_s^2
+2\sum_{r\in\{r<s\}_{+}}\gamma_r\gamma_s\bra{\psi_s}\psi_r\rangle|
C(\psi_s),
\\ |\gamma_{\overline{i}}|^2
C(\psi_{\overline{i}}),
\end{array}\right.
\end{equation}
where ${k\in \{k<l\}_{-}}$, ${s\in \{r<s\}_{+}}$ and
$\overline{i}\neq\{k,s\; |\; k\in\{k<l\}_{-},\;s\in\{r<s\}_{+}\}$.

\textit{Proof:} For the first inequality, we should proof that the
left-hand side is less than each term in the right-hand side. To do
so, we apply the Minkowski inequality (\ref{Minkowski}) several
times to Eq. (\ref{CGammaPrimeM}), followed by the fact that
$C^2(\psi_i,\psi_j)\le 1+|\bra{\psi_i}\psi_j\rangle|^2$. This proofs
the first line of  Eq. (\ref{Theorem8ArbitraryUpperM}). Now to reach
the second line of Eq. (\ref{Theorem8ArbitraryUpperM}), we first
partition the pairs $\{i<j\}$, with $1\le i< j \le m$, into two
distinct subsets of pairs $\{k<l\}_{-}$ and $\{r<s\}_{+}$. For the
first subset we expand $\ket{\psi_l}$ in terms of $\ket{\psi_k}$ and
$\ket{\psi_{k\perp}}$, i.e.
$\ket{\psi_l}=\bra{\psi_k}\psi_l\rangle\;
\ket{\psi_k}+\sqrt{1-|\bra{\psi_k}\psi_l\rangle|^2}\ket{\psi_{k\perp}}$,
and for the second one we expand  $\ket{\psi_r}$ in terms of
$\ket{\psi_s}$ and $\ket{\psi_{s\perp}}$, i.e.
$\ket{\psi_r}=\bra{\psi_s}\psi_r\rangle\;
\ket{\psi_s}+\sqrt{1-|\bra{\psi_s}\psi_r\rangle|^2}\ket{\psi_{s\perp}}$.
Putting these  into Eq. (\ref{CAlphaBetaPsiPhi}), we find
$C_{\alpha\beta}(\psi_k,\psi_l)=\bra{\psi_k}\psi_l\rangle^\ast
C_{\alpha\beta}(\psi_k)+\sqrt{1-|\bra{\psi_k}\psi_l\rangle|^2}C_{\alpha\beta}(\psi_k,\psi_{k\perp})$
and  $C_{\alpha\beta}(\psi_r,\psi_s)=\bra{\psi_s}\psi_r\rangle^\ast
C_{\alpha\beta}(\psi_s)+\sqrt{1-|\bra{\psi_s}\psi_r\rangle|^2}C_{\alpha\beta}(\psi_s,\psi_{s\perp})$,
respectively. Using these in Eq. (\ref{CGammaPrimeM}) and after some
calculations we find
\begin{eqnarray}\label{Theorem8ArbitraryUpperMProof}
& &\|\Gamma\|^2C(\Gamma^\prime)=
\{\sum_{\alpha\beta}|\sum_{\overline{i}}^{\prime}[(\gamma_{\overline{i}}^\ast)^2]
C_{\alpha\beta}(\psi_{\overline{i}}) \\ \nonumber
&+&\sum_{k\in\{k<l\}_{-}}[(\gamma_k^\ast)^2
+2\sum_{l\in\{k<l\}_{-}}\gamma_k^\ast\gamma_l^\ast\bra{\psi_k}\psi_l\rangle^\ast]
C_{\alpha\beta}(\psi_k)\\
\nonumber &+&\sum_{s\in\{r<s\}_{+}}[(\gamma_s^\ast)^2
+2\sum_{r\in\{r<s\}_{+}}\gamma_r^\ast\gamma_s^\ast\bra{\psi_s}\psi_r\rangle^\ast]
C_{\alpha\beta}(\psi_s)
\\  \nonumber
&+&2\sum_{i<j}(\gamma_i^\ast\gamma_j^\ast)\sqrt{1-|\bra{\psi_i}\psi_j\rangle|^2}C_{\alpha\beta}(\psi_i,\psi_{i\perp})|^2\}^{1/2},
\end{eqnarray}
where $\sum_{\overline{i}}^{\prime}$ denotes sum over all possible
values $1\le \overline{i}\le m$ such that $\overline{i}\neq\{k,s\;
|\; k\in\{k<l\}_{-},\;s\in\{r<s\}_{+}\}$; For example, for $m=3$ a
possible  partition for pairs $1\le i<j\le 3$ is $\{12\}_{-}$ and
$\{13,23\}_{+}$, therefore $\overline{i}=2$, but for the partition
$\{13\}_{-}$ and $\{12,23,\}_{+}$, $\overline{i}$ does not take any
value. Now by successive application of the Minkowski inequality,
followed by the fact that $C(\psi_i,\psi_{i\perp})\le 1$, we reach
to the second line of the inequality
(\ref{Theorem8ArbitraryUpperM}). This completes the proof for the
first inequality.

Next in order to proof the lower bound
(\ref{Theorem8ArbitraryLowerM}), we should proof that the left-hand
side is greater than each line in the right-hand side. For the first
line, we proceed just as that we done in the proof of the lower
bound of theorem 7, but using the fact that, here,
$C(\psi_i,\psi_j)\le \sqrt{1+|\bra{\psi_i\psi_j}\rangle|^2}$. On the
other hand to proof the second line of Eq.
(\ref{Theorem8ArbitraryLowerM}), we apply the inverse Minkowski
inequality to Eq. (\ref{Theorem8ArbitraryUpperMProof}) and separate
the first three terms from the last one, i.e.
\begin{eqnarray} \label{Theorem8ArbitraryLowerMProof}
& &\|\Gamma\|^2C(\Gamma^\prime)=
\{\sum_{\alpha\beta}|\sum_{\overline{i}}^{\prime}(\gamma_{\overline{i}}^\ast)^2
C_{\alpha\beta}(\psi_{\overline{i}}) \\ \nonumber
&+&\sum_{k\in\{k<l\}_{-}}(\gamma_k^\ast)^2
+2\sum_{l\in\{k<l\}_{-}}\gamma_k^\ast\gamma_l^\ast\bra{\psi_k}\psi_l\rangle^\ast
C_{\alpha\beta}(\psi_k) \\
\nonumber &+&\sum_{s\in\{r<s\}_{+}}(\gamma_s^\ast)^2
+2\sum_{r\in\{r<s\}_{+}}\gamma_r^\ast\gamma_s^\ast\bra{\psi_s}\psi_r\rangle^\ast
C_{\alpha\beta}(\psi_s)|^2\}^{1/2}
\\ \nonumber
&-&\{\sum_{\alpha\beta}|2\sum_{i<j}
(\gamma_i^\ast\gamma_j^\ast)\sqrt{1-|\bra{\psi_i}\psi_j\rangle|^2}C_{\alpha\beta}(\psi_i,\psi_{i\perp})|^2\}^{1/2}.
\end{eqnarray}
Now successive application of the inverse Minkowski inequality and
the Minkowski inequality  to the first and the second terms of the
above equation, respectively, followed by the fact that
$C(\psi_i,\psi_i\perp)\le 1$, leads to the lower bound
(\ref{Theorem8ArbitraryLowerM}).

\section{Conclusion}
We have used the notion of concurrence vector and have investigated
upper and lower bounds on the concurrence of the state as a function
of the concurrence of the superposed states. We have shown that the
amount of entanglement quantified by the concurrence vector is
exactly the same as that quantified by I-concurrence, so that we
could compare our results to those given in \cite{NisetPRA2007}. We
have shown that when the two superposed states are biorthogonal
states, the concurrence of the superposition can be written, simply,
as a function of the concurrence of  the superposed states. For
other situations such as one-sided orthogonal states, orthogonal
states, and arbitrary states, we have derived simple relations for
the upper and lower bounds of the superpositions. It is shown that
our bounds are, in general, tighter than the bounds obtained in Ref.
\cite{NisetPRA2007}. An extension of the results to the case with
more than two states in the superpositions is also given. It follows
from our results that the vectorial representation of the
generalized concurrence is more suitable in investigation of the
concurrence of superpositions.


\end{document}